# Voigt effect-based wide-field magneto-optical microscope integrated in a pump-probe experimental setup

T. Janda,[1,2,a)] L. Nádvorník,[3] J. Kuchařík,[1] D. Butkovičová,[1] E. Schmoranzerová,[1] F. Trojánek,[1] and P. Němec[1]

[1]*Faculty of Mathematics and Physics, Charles University, Ke Karlovu 3,121 16, Prague 2, Czech Republic*
[2]*Institute of Physics ASCR, v.v.i., Cukrovarnická 10, 162 53, Prague 6, Czech Republic*
[3]*Fritz-Haber Institute, Max-Planck Society, Faradayweg 4-6, 14195, Berlin, Germany*

**In this work we describe an experimental setup for spatially-resolved pump-probe experiment with integrated wide-field magneto-optical (MO) microscope. The MO microscope can be used to study ferromagnetic materials with both perpendicular-to-plane and in-plane magnetic anisotropy via polar Kerr and Voigt effects, respectively. The functionality of the Voigt effect-based microscope was tested using an in-plane magnetized ferromagnetic semiconductor (Ga,Mn)As. It was revealed that the presence of mechanical defects in the (Ga,Mn)As epilayer alters significantly the magnetic anisotropy in their proximity. The importance of MO experiments with simultaneous temporal and spatial resolutions was demonstrated using (Ga,Mn)As sample attached to a piezoelectric actuator, which produces a voltage-controlled strain. We observed a considerably different behavior in different parts of the sample that enabled us to identify sample parts where the epilayer magnetic anisotropy was significantly modified by a presence of the piezostressor and where it was not. Finally, we discuss the possible applicability of our experimental setup for the research of compensated antiferromagnets, where only MO effects even in magnetic moments are present.**

## I. INTRODUCTION

The investigation of electron spin-related phenomena is a rapidly evolving field. It is motivated by the fact that the classical charge-based electronics is approaching its physical limits as the circuit features are reaching nanometer scale.[1] For further substantial increase of computing performance new physical phenomena have to be employed in electronic devices. Among them the spin-based (or spintronic) devices represent very promising alternative.[2] Spin-sensitive magneto-optical (MO) spectroscopy has emerged as a particularly successful technique for the exploration of the electron spin dynamics, especially in semiconductors.[3] Here, various MO effects can be used as an efficient probe of magnetic order[4,5,6] and/or spin-polarized carriers[7]. The research aiming at a development of new spintronic devices inherently requires experimental techniques with at least a micrometer spatial resolution, which is a typical size of proof-of-principle devices.[8] This is relatively easy to achieve using MO microscopes that were developed for visualization of magnetic domains in the last decades.[9, 10,11]

---

[a)] Electronic mail: janda@karlov.mff.cuni.cz

Apart from the spatial resolution, the resolution in a time domain is needed as well to address the relevant time scales of the spin-related processes employed in the device operation.[12]

Many time-resolved studies of various materials were performed by the pump-probe experimental technique.[13] Here, a strong (pump) laser pulse modifies the investigated material and a time-delayed weaker (probe) laser pulse measures the temporal evolution of the pump-induced changes. The temporal resolution is given mainly by the duration of the laser pulses which is typically around 100 femtoseconds. To obtain *simultaneous* high temporal and spatial resolutions, the pump-probe experimental setup is typically combined with an optical microscope, which enables to visualize the sample topography (see also Fig. 1). In this setup, the imaging of the studied sample surface is achieved by its illumination using an independent (preferentially incoherent) light source and by detection of the reflected light intensity with a charge-coupled device (CCD) camera. The actual time-resolved measurements are performed using femtosecond laser pulses which are focused to a micrometer-sized spot by the microscope objective. Here, the achieved spatial resolution is given mainly by a numerical aperture of the used objective and by a wavelength of the laser light.[11] The probe laser pulses reflected back from the sample (or transmitted through it) are collimated again by the objective. Their intensity is measured by a photodetector as a function of a time delay between pump and probe pulses in order to obtain information about the photoexcited carrier dynamics.[14,15,16] Moreover, information about the magnetization or carrier spin dynamics can be also extracted if the pump-induced change of polarization of the probe pulses resulting from the sample MO activity is measured.[17,18,19] To assemble the spatially-resolved maps of MO response, either the sample is moved by a piezo scanning stage[17,19] or the incoming light beam is tilted with respect to the objective optical axis[18]. In any case, the measurement of extended sample areas with a sufficient spatial resolution is much more time-consuming task in these scanning MO microscopes[17,18,19,20] as compared to standard wide-field MO microscopes[11].

In this contribution we show that this serious drawback can be removed if the pump-probe experimental setup is incorporated in a magneto-optical microscope rather than in an optical microscope.

Our paper is organized as follows: First, we describe experimental details of the setup for pump-probe experiment with a high spatial resolution (part II.-a) and for wide-field MO microscope (part II-b.). Next we verify that domains in in-plane magnetized materials can be visualized using Voigt effect in this MO microscope. Finally, we test the overall performance of the combined setup by a study of ferromagnetic semiconductor (Ga,Mn)As attached to a piezoelectric actuator. This sample structure enables to achieve a strain-induced modification of magnetic anisotropy in (Ga,Mn)As, which is user-controlled by a voltage applied to the piezoelectric actuator [21,22,23], that is a very promising feature for many envisioned spintronic applications [20,24]. However, this functionality is achieved at the expense of rather inhomogeneous sample properties, which are caused by the inhomogeneous transfer of strain from the piezo-stack through the glue to the investigated material[25]. In particular, we illustrate that the wide-field MO microscope functionality of our experimental setup allowed us to identify sample regions with distinct effects of the piezo-induced strain, which are indistinguishable in normal optical microscope. Subsequently, we performed time-resolved MO measurements in such pre-selected areas and observed that their magnetization dynamics is also considerably different.

## II. EXPERIMENTAL SETUP FOR SPATIALLY-RESOLVED PUMP-PROBE EXPERIMENT WITH INTEGRATED MAGNETO-OPTICAL MICROSCOPE

The experimental setup shown in Fig. 1(a) consists of two parts which enable us to perform pump-probe experiments and magneto-optical imaging both independently and also in conjunction in a single combined experiment. In the following, we describe in detail both these functional parts.

### II-a. Pump-probe setup with high spatial resolution

As a light source for the pump-probe measurements we used Ti:sapphire laser (Spectra Physics, Mai Tai HP) generating ≈ 100 fs long linearly polarized laser pulses with a spectral width of ≈ 10 nm at a repetition rate of 80 MHz. The central wavelength of the laser pulses can be continuously tuned between 690 and 1040 nm. Each laser pulse is split into strong pump pulse and weaker probe pulse by a polarizing beamsplitter (PBS) and the division ratio can be smoothly tuned by a half-wave plate (HWP) placed in front of it. In addition to that, a combination of a HWP and a polarizer in each arm of the setup allows for setting the required intensities of the pump and probe beams separately. Intensity of the pump beam is modulated at a frequency of ≈ 2 kHz by an optical chopper and pump pulses are time-delayed in the optical delay line. To spectrally separate the pump and probe pulses, they pass through different dielectric edge filters F1 and F2, which transmit only the short-wavelength part and long-wavelength part of the original laser pulse, respectively.[26] This prevents scattered pump photons from reaching the detectors (D), which are placed behind additional filters F2 [see Fig. 1(a)], and, consequently, it reduces the noise in the measured curves considerably. After spectral filtering, the pump pulses are reflected towards the sample by a piezo-controlled mirror (Newport, AG-M100D) which sets the angle of the pump pulse incidence on the objective lens and, consequently, the pump laser spot position on the sample surface. Both pump and probe beams pass a series of polarization components (polarizer, half- and quarter-wave plate (QWP)) used to set the desired polarization state – linear polarization in the case of the probe beam and circular or linear polarization in the case of the pump beam. The beams are merged by a beamsplitter (BS) and they are focused by a microscope objective (Mitutoyo Plan Apo 20x, with numerical aperture (NA) of 0.42) to ≈ 1 μm spots on the sample surface. The sample is placed in an optical cryostat (Advanced Research Systems), where the temperature can be changed from 10 to 800 K. The cryostat is mounted between the poles of an electromagnet (Walker Scientific, model HV-4H) generating magnetic field up to 600 mT. The objective is attached to a 3D piezo-positioner (Newport, non-magnetic version of NPXYZ100SG-D), which allows for precise focusing and choosing of the area of interest on the sample surface. The achievable spatial separation of pump and probe spots on the sample surface is limited by the size of the objective input aperture – in our case, their separation can exceed 20 μm.[27]

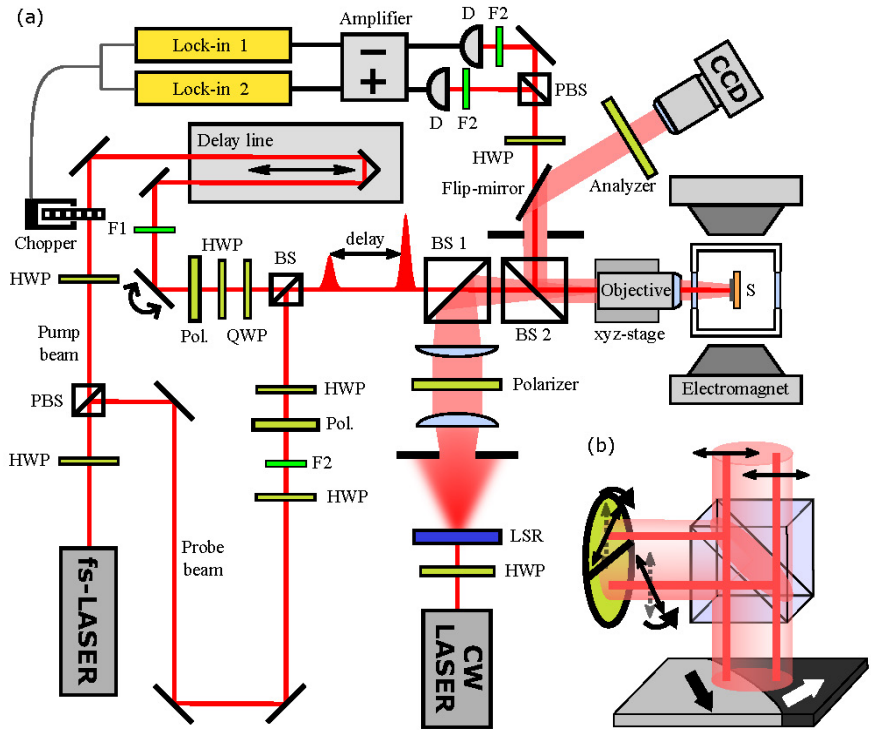

**FIG. 1. Pump-probe experimental setup with a high spatial resolution combined with a wide-field MO microscope.** (a) Sketch of the experimental layout. Pump-probe experiment: Each pulse emitted by femtosecond (fs) laser is divided into pump and probe pulse by a polarizing beamsplitter (PBS) with an intensity ratio set by a half-wave plate (HWP). Another pair of HWP and polarizer is used to set the intensity in each beam. Filters F1 and F2 are used to spectrally separate the pump and probe beams. Sets of polarizer, half- and quarter-wave plate (QWP) define the polarization of pump and probe pulses which are then merged by a beamsplitter (BS), with a variable mutual time delay set by a delay line, and focused on the sample (S) by an objective lens. The sample is placed in an optical cryostat between the poles of an electromagnet. Beams reflected from the sample are directed by BS 2 towards an optical-bridge detection system where the pump pulses are suppressed by filters F2. MO microscope: HWP and polarizer are used to set intensity and vertical (*s*) polarization of light emitted by continuous (cw) laser. Laser speckle reducer (LSR) suppresses interference effects due to the laser coherence. The optical system creates nearly parallel beam illuminating the sample and reflected light is guided by an optional flip-mirror through an analyzer to a CCD camera. (b) Working principle of a wide-field MO microscope. Polarization plane of light reflected from places with different orientation of magnetization in the sample plane is rotated by a different angle resulting in a different intensity transmitted through the analyzer.

The reflected probe pulses are collimated by the objective lens and propagate, with the flip-mirror lowered, towards the detection part of the setup – so called optical bridge. Here the probe beam is split by a polarizing beamsplitter in two parts detected by individual detectors. The corresponding electrical signals are then added (for the transient reflectivity measurement) and subtracted (for the MO signal measurement), amplified and processed by lock-in amplifiers, which extract the signal modulated at the chopper-frequency. Before the actual measurement, the bridge has to be balanced with the pump beam blocked. This is accomplished by rotating the HWP in front of the beamsplitter until the difference signal is zero, which represents a reference state without the sample excitation. When the sample is excited by the pump pulse, its magnetic properties change. This leads (via MO effects) to a change of probe polarization and, consequently, it shifts the bridge off balance. Therefore, the measured difference signal is proportional to the magnetic changes in the sample. For example, this setup can be very efficiently used for a study of spin diffusion and drift, as we demonstrated in undoped GaAs/AlGaAs epilayers recently.[27]

**II-b. Wide-field magneto-optical microscope**

Polarizing microscopy is one of the oldest, simplest and most powerful techniques used for investigation of magnetic domain structure in ferromagnetic materials. It utilizes magneto-optical (MO) effects which cause change in polarization state of light transmitted through or reflected from a magnetically ordered material.[9,10,11] As most of the magnetic media and/or the substrates used for their preparation are not transparent in visible and near-IR wavelength range, this technique is typically implemented in the reflection geometry.[10] The most straightforward is a visualization of magnetic domains in materials with a perpendicular-to-plane magnetic anisotropy where the polar Kerr effect (PKE) can be used. Here, a linearly polarized light experiences a change of polarization upon reflection on a magnetized medium and the magnitude of this polarization change is proportional to the projection of the magnetization along the light propagation. Consequently, the microscopes based on PKE use usually a normal incidence of light and they allow us to distinguish between magnetic domains with opposite magnetization directions.[28] The visualization of magnetic domains with in-plane magnetic anisotropy is considerably more complicated. One possibility is to use the longitudinal or transversal Kerr effects [29,30] but it requires a large angle of incidence.[11] The whole family of Kerr effects belongs to a broader class of MO effects that are linear in magnetization, i.e., their characteristic feature is a polarization rotation sign change if the magnetization direction in the ferromagnetic material is flipped. These effects are typically larger than the MO effects that are quadratic in magnetization, they are simpler to separate from magnetization-independent optical signals and, therefore, they are more commonly implemented.[31] However, quadratic MO effects, like Voigt (or Cotton-Mouton) effect, can be also used for a research of ferromagnetic materials, in particular, of those with the in-plane magnetic anisotropy.[4,5,31] Moreover, the quadratic MO effects have an additional advantage that they do not vanish in certain situations where the linear MO effects do. For example, in compensated antiferromagnets the quadratic MO effects are present because contributions from the two spin sublattices with opposite spin orientations do not cancel.[32] We discuss the potential of quadratic MO effects for a research of antiferromagnets in the concluding section of this paper.

A layout of the wide-field MO microscope, which is integrated with the pump-probe setup, is shown in Fig. 1(a). (Note that we use the term "wide-field" MO microscope throughout the paper just to clearly distinguish it from the "scanning" form of MO microscope.) As a light source for the microscope, we used continuous-wave Ti:sapphire laser (Spectra Physics, model 3900S), which has a gap-free tunability from 700 to 1000 nm. The possibility to change the illumination wavelength is very convenient for a fine tuning of the microscope performance for a given sample, as we show in part III, and also for a utilization of the microscope for a research of various material systems with MO spectra located in different spectral regions. The disadvantage of using coherent laser light for imaging is the presence of laser speckles. However, these interference effects can be considerably suppressed by a used laser speckle reducer (Optotune, model LSR-3005-24D-VIS), where one diffuser is periodically moved at a high frequency relative to another static diffuser. The desired linear polarization of light is set by a polarizer (Thorlabs, LPNIR 100-MP) and the illumination intensity can be continuously adjusted in a broad range by a half-wave plate (HWP) placed in front of the polarizer. The beam is then focused to the back focal plane of the microscope objective which creates a nearly parallel beam illuminating the sample. Light reflected from the sample passes through an analyzer and is detected by a high speed CCD camera (Allied Vision Tech, model Prosilica GX 1050).

Our MO microscope is sensitive to a perpendicular-to-plane projection of magnetization via PKE and also to an in-plane projection of magnetization thanks to Voigt effect (VE). Moreover, both these MO effects generate the maximum signal at normal incidence of light on the sample. Therefore, this MO microscope can visualize magnetic domains both in samples with perpendicular-to-plane and in-plane magnetization orientation without any change in its settings. As we do not need large angles of incidence we could use an objective lens with a relatively low numerical aperture (NA = 0.42). Compared to large-NA objectives, it has many advantages like lower cost and larger working distance. However, the major advantage of low-NA objectives for our combined optical setup is that they usually have relatively large input aperture which allows us to reach larger distance between the pump and probe spots on the sample, as discussed in the previous chapter. This is necessary to perform the spatially resolved pump-probe measurements.[27]

The basic working principle of a MO microscope is shown in Fig. 1(b). Polarization plane of a linearly polarized light reflected from regions with a different orientation of magnetization is rotated by a different angle (via MO effects) which results in different intensity transmitted through the analyzer. Magnetic domains with different magnetization directions thus have different brightness in the image captured by the CCD camera.[11] The linear polarization of light used to illuminate the sample in our microscope can be chosen either vertical (*s*-polarization) or horizontal (*p*-polarization). For other orientations of the incident linear polarization the beamsplitters cause a significant phase shift between the *s*- and *p*-polarized component. This results in a strong light ellipticity, which substantially decreases the contrast observable between the domains. Following the idea of Fig. 1(b), for an appropriate orientation of the analyzer one should be able to totally block the light reflected from places with one orientation of magnetization whereas the light reflected from places with a different orientation of magnetization should be partially transmitted. In reality, due to depolarization effects on the optical surfaces in the microscope, for any orientation of the analyzer there is always light coming through it that is creating a non-magnetic background in the image.[33] This undesirable contribution can be removed by a subtraction of a reference image taken in a state with a fixed magnetization orientation, e.g., in the magnetically saturated state. This subtraction procedure highlights the changes in magnetization orientation with respect to the reference state and, consequently, it enables to detect magnetization-induced polarization rotations of the order of 100 μrad. The visibility of the domain pattern in images taken by the microscope can be characterized by a magneto-optical contrast defined as

$$C_{MO} = \frac{I_2 - I_1}{I_2 + I_1 + 2I_B} \times 100\%, \qquad (1)$$

where $I_1$ and $I_2$ stand for the intensities extracted from the taken image after a subtraction of the reference image in two magnetic domains with different magnetization orientations. $I_B$ stands for the "typical" background intensity of the reference image (e. g., extracted from the middle of the reference image). The observed MO contrast is mainly affected by an orientation of the analyzer. In order to reach the maximum $C_{MO}$ the analyzer should be rotated 3° – 5° from the direction of maximum extinction with the polarizer.[34]

## III. VERIFICATION OF VOIGT EFFECT-BASED MO MICROSCOPE FUNCTIONALITY

As discussed in the previous chapter, our MO microscope is in principle sensitive both to a perpendicular-to-plane projection of magnetization via the PKE and to an in-plane projection of magnetization by the VE. In reality, it is considerably more difficult to visualize magnetic domains with an in-plane magnetization orientation. Therefore, we will focus on the latter case in the following. The demonstration of the MO microscope functionality was performed using a 20 nm thick epitaxial layer of $Ga_{1-x}Mn_xAs$ on GaAs substrate. (Ga,Mn)As is a diluted ferromagnetic semiconductor[35,36] with an in-plane magnetic anisotropy that is characterized by two easy axes (EAs) making an angle $\xi$ (see Fig. 2(a) and Appendix). If an external magnetic field $H_{EXT}$ is applied in the sample plane at an angle with respect to both EAs (see Appendix), the magnetization position switches between the two EAs [e.g., no. 1 and 2 in Fig. 2(a)] and the corresponding MO contrast appears. This MO contrast is proportional to a change in the VE-induced rotation $\Delta\beta$ of the light polarization plane that is described by a formula[37]

$$\Delta\beta(EA1) - \Delta\beta(EA2) = 2P^{VE} \cos 2(\gamma - \beta) \sin \xi \quad . \tag{2}$$

Here $P^{VE}$ is the (spectrally dependent) MO coefficient describing the strength of VE and angles $\gamma$ and $\beta$, both measured from the [100] crystallographic direction, denote the orientation of the easy axes bisector and the illuminating light polarization plane orientation, respectively (see Fig. 4 in Ref. 37 for a depiction of all the angles). It is immediately apparent from Eq. (2) that to maximize the MO contrast resulting from the magnetization switching from one EA to the other EA, the orientation of the light polarization should be chosen along (or perpendicular to) the easy axes bisector, as illustrated in Fig. 2(a). Moreover, the maximum MO contrast is obtained in a case of mutually perpendicular EAs ($\xi$ = 90°). In our test sample with Mn content of $x \approx 0.03$ the EAs are tilted for ≈ 15° towards the [-110] crystallographic direction [see Fig. 2(a)]. Consequently, the angle between the EAs is $\xi \approx 60°$. All the experiments were performed at temperature of 15 K, which is considerably lower than the sample Curie temperature of 77 K.[38]

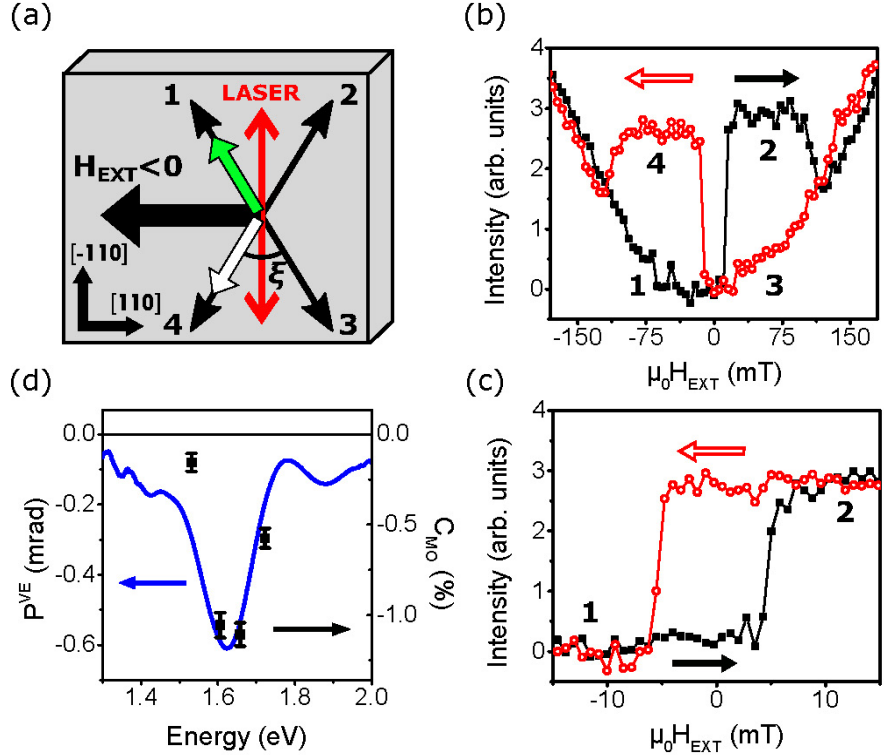

**FIG. 2. Magneto-optical properties of the test (Ga,Mn)As sample**. (a) Implemented experimental geometry. Thin black arrows (numbered 1 – 4) represent magnetization easy axis directions in the sample. Laser light polarization was set along the [-110] crystallographic direction in the sample. Before the hysteresis loops measurements the sample was initiated by an application of saturating external magnetic field $\mu_0 H_{EXT}$ = -600 mT along the [-1-10] direction. Green arrow along the direction "1" represents a magnetization state in the majority of the sample when the external magnetic field is reduced to zero. However, in certain parts of the sample the magnetization is oriented along the direction "4", as indicated by the white arrow (see text for details). (b) and (c) Major and minor hysteresis loops, respectively, measured by the MO microscope via Voigt effect. Numbers represent the magnetization states depicted in part (a) and arrows depict the direction of the magnetic field change. (d) Independently measured spectral dependence of the Voigt effect magnitude (line) and MO contrast $C_{MO}$ (points) calculated from images taken by the MO microscope using Eq. (1).

To confirm that the MO contrast observed in the microscope really originates from the VE, we first deduced MO hysteresis loops from the measured images. To do so, we selected a place on the sample where no domain structure was visible at zero magnetic field, i.e., this area was in a single-domain state. During the measurement, $H_{EXT}$ was changed in small steps starting from large negative values across zero to large positive values and back with the microscope taking a picture of the sample in each step (without any subtraction of the reference image). By averaging the gray scale intensity over the whole image at each field value we obtained a dependence of the detected intensity on the magnetic field – a hysteresis loop. The measurement was repeated 6 times and the obtained curves were averaged and shifted vertically to zero (to remove the non-magnetic background in the images). In Fig. 2(b) we show the hysteresis loop measured for a magnetic field applied along the [-1-10] and [110] crystallographic directions in the sample plane in a large range of fields. The obtained loop is even in the magnetic field which immediately reveals that the responsible MO effect is even in magnetization, as it should be in the case of VE. Moreover, the hysteresis loop is quite similar to those measured previously via this MO effect in an analogous sample (see Fig. 1 in Ref. 39). In fact, by the inspection of the measured data we can identify the corresponding magnetization orientations in the sample during the field sweeps.[39] The hysteresis loop shown in Fig. 2(b) represents so called major hysteresis loop that addresses all four possible magnetization states ($1 \rightarrow 2 \rightarrow 3 \rightarrow 4 \rightarrow 1$) when going from a strong negative $H_{EXT}$ to a strong positive $H_{EXT}$ and backward. Theoretically, in the case of an application of $H_{EXT}$ close to the highly symmetrical direction [-1-10], both magnetization directions 1 and 4 could be populated when $H_{EXT}$ is reduced from strong negative values to zero. However, we observed experimentally a single domain state in the direction 1 only, which is probably a consequence of a slight misalignment of $H_{EXT}$ from the direction [-1-10] towards the direction [-110]. (The zero-magnetic-field-state assignment to the EA 1 was verified by an additional experiment, which is not shown here, when $H_{EXT}$ was applied in the direction tilted from the direction [-1-10] towards the direction [1-10] by $\approx 6°$, which favors the EA 4 instead.) For a strong negative $H_{EXT}$ the magnetization no longer points to the direction 1 (or 4) but it is tilted towards the applied field direction. This field-induced magnetization tilt is partially responsible for the slope observed in the measured data for the fields stronger than $\approx 100$ mT. However, a significant part of this slope is probably caused by the field-induced shift of the objective lens with respect to the sample plane in strong fields, which slightly varies the amount of light reaching the CCD camera. In Fig. 2(c) we show an upper minor hysteresis loop[39] obtained when sweeping weak $H_{EXT}$, which addresses only two closest magnetization states ($1 \rightarrow 2 \rightarrow 1$).

In an additional control experiment we studied the spectral dependence of the MO contrast. The line in Fig. 2(d) is an independently measured spectrum of the VE-induced polarization rotation in the studied sample.[31,37] The points represent the MO contrast calculated using Eq. (1) from the images captured by the MO microscope at four different photon energies. As expected, both quantities show very similar spectral dependence which further confirms the VE as the responsible MO effect in our microscope. (The deviation at the lowest energy is probably caused by a decreased quantum efficiency of the CCD camera in this spectral range, which was affecting the observed MO contrast.) As the MO contrast is the largest around 1.60 eV, we used this photon energy for the rest of the experiments performed on this sample.

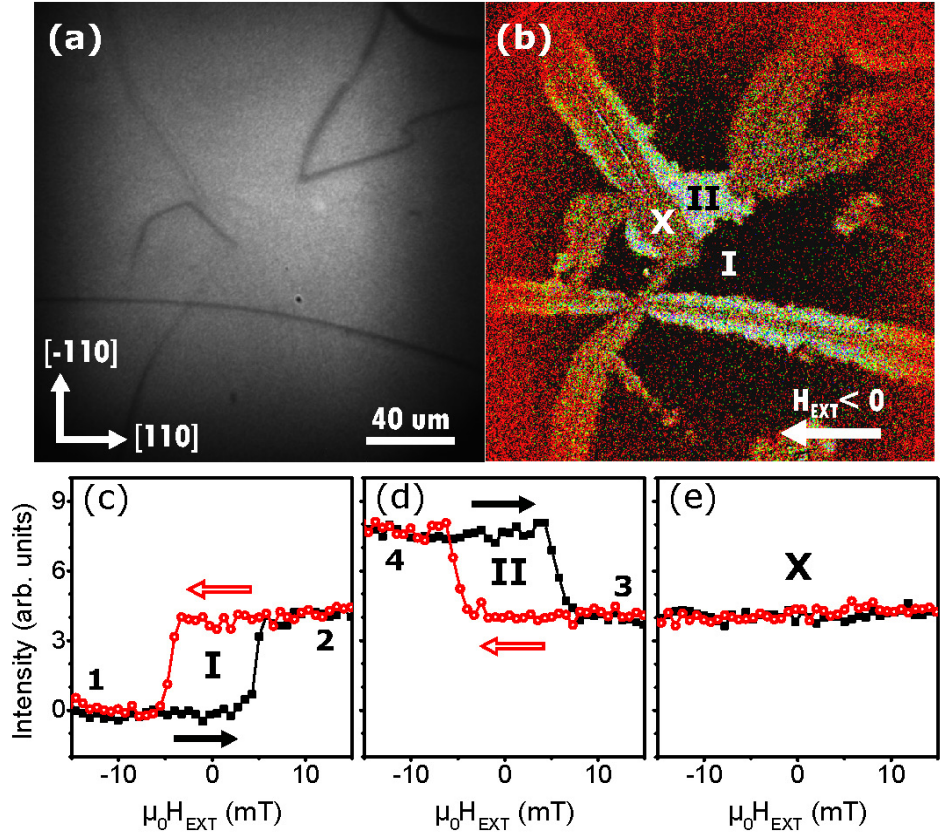

**FIG. 3. Imaging of the sample around the surface imperfections**. (a) Image showing the position of several defects on the sample surface. (b) MO image of the same area at -25 mT (with a reference image taken at 25 mT). (c), (d) and (e) Minor hysteresis loops measured in places marked as I, II and X in (b). Numbers represent the magnetization states depicted in Fig. 2(a).

In a majority of the studied (Ga,Mn)As sample the domain structure observation was complicated by the high velocity of the magnetization switching process. Therefore, it was easier to perform the measurement closer to the sample edge where structural defects acting as pinning sites were located and, consequently, where the magnetic field-induced domain wall motion was slower. In Fig. 3(a) we show an image of the sample where several scratches are present on the sample surface. Fig. 3(b) shows an image of the same area with highlighted changes in light polarization orientation upon reflection from the sample, which was obtained by the following procedure. First, a reference image was taken with 25 mT field applied, then the field was switched to -25 mT and the second image was taken. Finally, both images were subtracted which highlighted the field-induced changes of the magnetization orientation with respect to the situation at 25 mT. We observed that the magnetic domain patterns clearly followed the structural defects distribution on the sample surface. In fact, three different intensity levels can be found in the image that are marked as *I*, *II*, and *X* in Fig. 3(b). To illustrate the difference between these three regions more clearly, we show the minor hysteresis loops extracted from these image parts in Figs. 3(c), (d) and (e), respectively. Far away from the scratches – in the dark parts labeled as I in Fig. 3(b) – (Ga,Mn)As behaves exactly the same as in the sample parts without visible defects [c.f., Fig. 2(c) and Fig. 3(c)], i.e., the magnetization is switching between the states 1 and 2, as discussed before (with corresponding switching fields $\mu_0H_{12}$ = -$\mu_0H_{21}$ = 4.6 ± 0.2 mT). On the other hand, closer to the scratches – in lighter regions labelled *II* in Fig. 3(b) – the switching occurs between the states 4 and 3, as depicted by a white arrow in Fig. 2(a) (with switching fields $\mu_0H_{43}$ = -$\mu_0H_{34}$ = 5.4 ± 0.2 mT). The most probable explanation of why the magnetization state 4 was populated in regions *II* when $H_{EXT}$ was reduced from large negative values to zero (unlike in regions *I* where the state 1 was populated), is the strain modification of the (Ga,Mn)As layer induced by the presence of scratches. As we describe in detail in the Appendix, this strain modification can locally rotate the EAs orientation for a few degrees that could be sufficient for a preferential population of the state 4 in our case when $H_{EXT}$ was applied very close to the highly symmetrical crystallographic direction [-1-10] [see Fig. 2(a)]. Finally, in areas directly adjacent to the scratches, which are labelled *X* in Fig. 3(b), no change in the detected light intensity was observed in the studied field range. The most probable explanation of this behavior is that the 20 nm thick (Ga,Mn)As film was missing in these sample parts due to the mechanical damage. Alternatively, the magnetic film could be still present even around the scratches but its magnetic anisotropy might be so severely altered by the strain modification that the magnetization was not switched in the investigated range of $H_{EXT}$ (see Appendix). To show the domain pattern evolution during the field sweep in a more illustrative way we performed also a separate measurement with a smaller field-step and assembled a video from the taken images (see Supplementary material).

## IV. INVESTIGATION OF PIEZO-INDUCED STRAIN INHOMOGENEITY BY COMBINED MO IMAGING AND PUMP-PROBE EXPERIMENT

Modification of magnetic anisotropy and manipulation with spins via a controlled strain application belongs to very promising ways towards spintronic applications.[20,21,23,24,40] There are several ways how the strain control can be achieved among which the piezo actuators are especially interesting because they enable a voltage-control of the strain in a wide temperature range.[25,41] This technology was particularly successful in (Ga,Mn)As where it enabled *in situ* control of magnetic anisotropy and, consequently, the discovery of several new physical phenomena.[21,22,23] The commercially available piezo actuators produce strain of about $5 \times 10^{-4}$ for the applied voltage of 150 V.[25,41] However, besides this voltage-tunable strain there is also a contribution arising from a difference in the sample/stressor thermal contraction at cryogenic temperatures, which can be even stronger than that induced in the sample by the applied bias.[21] Moreover, the strain distribution in the sample can be rather inhomogeneous due to the non-ideal transfer of the strain from the piezo-stack through the glue to the investigated material[25] that can result in position-dependent magnetic properties in the sample. Unlike magnetic anisotropy, optical properties of the sample (e.g., its reflectivity) are not modified by the strain of this magnitude and, consequently, no effects of strain can be seen in an optical microscope. However, as we illustrate below, they can be very efficiently detected by a magneto-optical microscope.

To test this, we used a sample containing 20 nm thick $Ga_{1-x}Mn_xAs$ epilayer with $x \approx 0.038$ ($\xi \approx 53°$, $T_C = 96$ K) grown on GaAs substrate that was polished down to 200 μm and attached by an Epotek glue to the uniaxial piezostressor (Piezomechanik GmbH, part no. PSt 150/2×3/7).[25] All experiments were performed at temperature of 15 K with the voltage of ±150 V applied to the piezostressor. Both the fs-laser (for the pump-probe experiment) and cw-laser (for the MO imaging) were tuned to 774 nm to maximize the MO signal due to VE in this epilayer.[31] In Fig. 4(a) we show an optical image of one sample part with depicted crystallographic directions in (Ga,Mn)As, positions of EAs (in a bare epilayer without the piezostressor), and directions of $H_{EXT}$ and the bias-induced strain. Apart from a crack there is no other structure visible on the sample surface. However, the sample inspection with the MO microscope revealed a presence of two regions, located left and right from the crack, which have very different behavior in external magnetic field – see Figs. 4(b) - (d). While on the right side we do not register any change of MO contrast with $H_{EXT}$, on the left side there is a clearly visible modification of MO contrast, which is reflecting the process of magnetization switching via a domain wall motion. To make this different behavior more apparent, in Fig. 5(a) we show hysteresis loops observed in the sample parts labelled as A and B in Fig. 4(a). If no strain is present in the sample, application of $H_{EXT}$ at a direction close to the EA leads to a switching of magnetization by 180° (see Fig. 6(b) in Appendix). However, this magnetization switching could not be visualized by the MO Voigt effect, which is even in magnetization. Consequently, the presence of the MO signal change with $H_{EXT}$ – and, in particular, the presence of the magnetization switching *before* the direction of $H_{EXT}$ is reversed – is a clear signature of the strain-induced change of magnetic anisotropy (see Figs. 6(c) and (d) and the adjacent discussion in Appendix). Consequently, these MO images enabled us to identify sample parts where the epilayer magnetic anisotropy was significantly modified by a presence of the piezostressor (part A) and where it was not (part B).

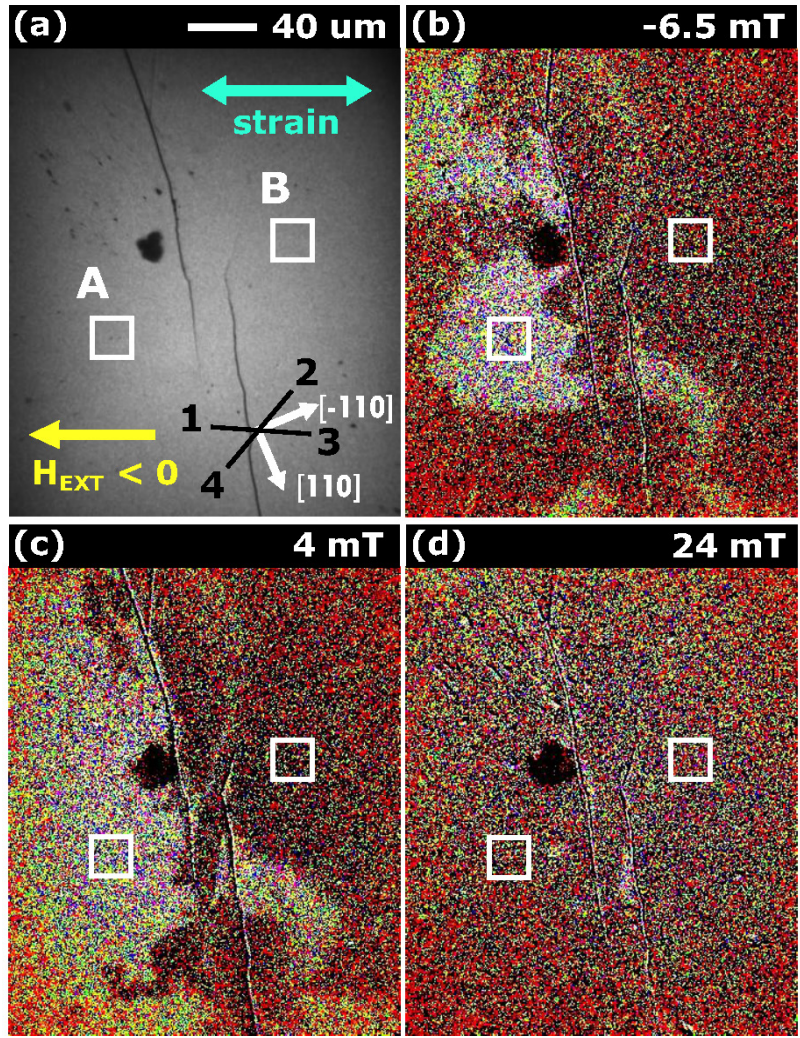

**FIG. 4. Investigation of magnetic homogeneity of the (Ga,Mn)As sample attached to a piezostressor by the MO microscope**. (a) Image of an area selected for a visualization of the differences in magnetic properties in different parts of the sample. Black lines marked "1" to "4" depict the EAs orientation in bare (Ga,Mn)As epilayer. The directions of applied magnetic field $H_{EXT}$ and bias-dependent strain induced by the piezostressor are marked by arrows. The squares marked A and B show the regions chosen for subsequent hysteresis and pump-probe measurements. (b) – (d) Images obtained by the MO microscope at different applied magnetic fields showing the magnetic inhomogeneity of the sample; the reference image was taken at -30 mT. All these images were measured at +150 V applied to the piezostressor and no significant change with the piezo-voltage was observed.

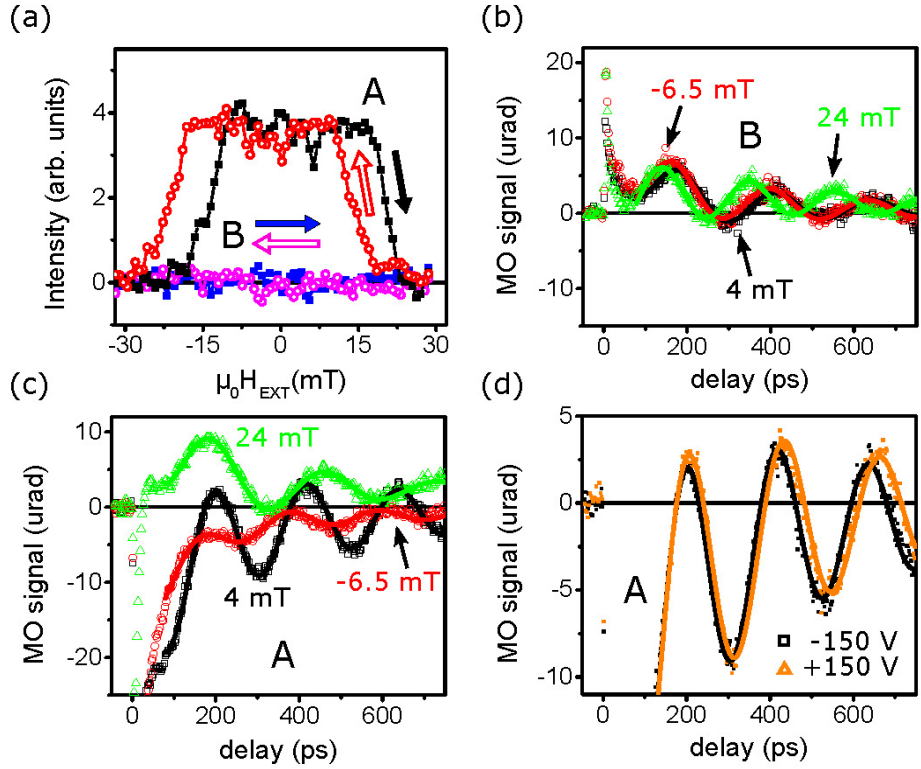

**FIG. 5. Comparison of magnetic behavior observed in the regions A and B depicted in Fig. 4(a)**. (a) Magnetic hysteresis loops extracted by the MO microscope from both regions. (b) and (c) Measured time-resolved oscillatory MO signal due to the magnetization precession induced by an impact of pump pulse for three different values of $H_{EXT}$ in regions B and A, respectively (points). Lines are fits by damped harmonic function. (d) Change of oscillatory signal induced in region A by applying ± 150 V to the piezostressor for $H_{EXT}$ = 4 mT. No bias dependence was observed in region B.

In Figs. 5(b) and (c) we demonstrate that the different magnetic properties of the regions A and B are accompanied also by significant differences in ultrafast magnetization dynamics observed by the pump-probe technique. In this experiment, absorption of the pump laser pulse leads to a photo-injection of electron-hole pairs. The subsequent fast nonradiative recombination of the photo-injected electrons induces a transient increase of the lattice temperature (within tens of picoseconds after the impact of the pump pulse). The laser-induced change of the lattice temperature then leads to a change of the easy axis position.[22] As a result, the magnetization starts to follow the easy axis shift by a precessional motion with a frequency that is directly connected with the sample magnetic anisotropy.[38] On a longer time scale, a dissipation of the heat leads to the return of the easy axis to the equilibrium position and the precession of magnetization is stopped by Gilbert damping. In the region B [see Fig. 5(b)], there is almost no dependence of the magnetization precession on $H_{EXT}$ except for a slight increase of the precessional frequency. This is consistent with the expected weak modification of the magnetic energy density minimum width with $H_{EXT}$ applied close to the EA direction (see Fig. 6(b) in Appendix). Moreover, we have not observed any dependence of the measured dynamical MO signal on the voltage applied to the piezostressor which further confirms that the region B is not properly attached to the piezostressor. On the other hand, in the region A [see Fig. 5(c)] there is a rather strong dependence of the precession amplitude, frequency and phase on $H_{EXT}$ which is in accord with the expected large change of the strain-modified magnetic anisotropy with $H_{EXT}$ (see Fig. 6(d) in Appendix). The precession of magnetization in the region A can be also modified by a voltage applied to the pizostressor [see Fig. 5(d)] but this change is rather small, presumably due to a stronger influence of the sample/stressor thermal contraction at cryogenic temperatures relative to the voltage-tunable strain.[21]

## IV. CONCLUSION AND OUTLOOK

We have constructed an optical pump-probe setup for time-resolved experiments with a high spatial resolution that is supplemented with a wide-field magneto-optical (MO) microscope, which can be used to study not only the sample topography but also the magnetization domain pattern on the sample surface. Our microscope employs geometry with a normal incidence of light on the sample where perpendicular-to-plane and in-plane magnetized domains can be visualized using polar Kerr and Voigt effects, respectively. The functionality of the whole combined experimental setup was tested on an in-plane ferromagnetic semiconductor (Ga,Mn)As. It was revealed that the presence of mechanical defects in the (Ga,Mn)As epilayer can significantly alter the magnetic anisotropy in their proximity. It was also demonstrated that this experimental setup is ideally suited for a rapid inspection of spatially-inhomogeneous samples which can be followed by a dynamical pump-probe magneto-optical experiments in the pre-selected sample parts.

Finally, we would like to briefly discuss the applicability of our experimental setup to other material systems. As already mentioned in Chapter II-b., even though the existence of MO effects quadratic in magnetization (e.g., the Voigt effect) has been well known for a long time they are only rarely used for MO imaging in ferromagnetic materials.[11] The main reason is that there exist other – and usually considerably stronger – MO effects, which are linear in magnetization, that can be utilized for this purpose. In particular, longitudinal or transversal Kerr effects are frequently used for a visualization of in-plane domains in ferromagnets (FMs).[11] However, the situation is markedly different in another type of magnetically ordered materials – in antiferromagnets (AFs). Apart from being used as a reference layer in spin valves, where they fix the magnetization orientation in FM due to the exchange bias,[42] AFs were usually overlooked from the point of view of practical applications. However, AFs have many appealing features which are intended to be used in the recently proposed concept of antiferromagnetic spintronics.[43] In particular, antiparallel spin sublattices in AFs, which produce zero dipolar fields, imply the insensitivity to magnetic-field perturbations and multi-level stability in magnetic memories. Moreover, spin dynamics in AFs is orders of magnitude faster than in FMs[13] which could eventually lead to a realization of electronical devices working at THz frequencies.[43] On the other hand, the absence of a net magnetic moment and the ultra-fast magnetization dynamics make AFs notoriously difficult to study by common magnetometers or magnetic resonance techniques. Especially in the case of epitaxial thin films of nanometer thickness, which form the building blocks of spintronic devices, the portfolio of experimental techniques that can be used for their investigation is very limited and, moreover, it typically requires utilization of large scale facilities.[44] Therefore, detection techniques based on the interaction of electromagnetic radiation with AFs attract significant attention nowadays.[32] From these techniques, MO studies represent the most evolved group which, moreover, enables to achieve both the spatial and temporal resolutions. For fully compensated AFs, the signals from oppositely oriented magnetic sublattices cancel each other for MO effects linear in magnetization which leaves only the MO effects quadratic in magnetization as suitable probes for these materials.[32] Very recently, we have demonstrated the potential of MO Voigt effect for a study of thin film of antiferromagnetic metal CuMnAs,[44] which is the prominent material used in the first experimental realization of room-temperature antiferromagnetic memory chips.[45] From this point of view, the table-top Voigt effect-based MO microscope represents a rather interesting tool because, in principle, it might be used also for a visualization of antiferromagnetic domains. Nevertheless, the insensitivity of AFs to moderately strong external magnetic fields complicates considerably the experimental separation of magnetic order-related MO signal from other sources of the light polarization change (e.g., strain- or crystal structure-related). Eventually, this separation can be accomplished by using the temperature and/or spectral dependence of the Voigt effect, which is markedly different from that of, e.g., strain-induced polarization changes. However, the actual utilization of the Voigt effect-based MO microscope for the study of AFs is yet to be demonstrated and only future will show its real research potential for the investigation of magnetically ordered materials with zero net moment.

**SUPPLEMENTARY MATERIAL**

The supplementary material contains a video showing the domain pattern evolution in the (Ga,Mn)As sample studied in Chapter III during a magnetic field sweep. The video was assembled from individual images taken by the MO microscope at the depicted fields when the field-induced changes were stabilized in the sample.

**ACKNOWLEDGMENTS**

The authors are indebted to V. Novák for a preparation of the (Ga,Mn)As samples. This work was supported by the Grant Agency of the Czech Republic under Grant No. 14-37427G and by EU FET Open RIA Grant no. 766566.

**APPENDIX: MAGNETIC ENERGY DENSITY CALCULATIONS**

In the case of our test material, $Ga_{1-x}Mn_xAs$ thin epitaxial layer, the in-plane magnetic anisotropy is a result of a competition between the cubic anisotropy, where the minima of the functional of magnetic energy density are along the [100] and [010] crystallographic directions, and the uniaxial in-plane anisotropy, with the minimum along the [-110] direction.[38] At a given temperature, the relative magnitude of these two contributions is strongly influenced by a concentration $x$ of Mn substituting Ga in the host lattice: the cubic anisotropy decreases with $x$ while the uniaxial anisotropy is nearly independent of $x$ (see Fig. 4(a) in Ref. 38). Consequently, the in-plane magnetic anisotropy of this material can be smoothly tuned by changing the value of $x$. In a presence of a mechanical stress the sample magnetic anisotropy is modified by an additional uniaxial contribution.[46] The magnetization switching process during an external magnetic field sweep can be understood through a minimization of the total magnetic energy density $E_{TOT}$, which has the following dependence[21,46] on the in-plane magnetization orientation $\varphi$:

$$E_{TOT} = \frac{1}{2}M_S \left[ \frac{1}{4}H_C \sin^2 2\varphi - \frac{1}{2}H_U(1 - \sin 2\varphi) - H_S \cos^2(\varphi - \delta) - H_{EXT}\cos(\varphi - \varphi_H) \right]. \quad (S1)$$

Here $M_S$ is the saturation magnetization, $H_C$, $H_U$ and $H_S$ are the effective anisotropy fields arising from cubic, uniaxial and strain-induced contributions to the total magnetic anisotropy, respectively, and $H_{EXT}$ is the external magnetic field magnitude. Angle $\varphi_H$ denotes the magnetic field direction and $\delta$ is the angle at which the strain-induced energy minimum is located (all angles are measured with respect to the [100] in-plane crystallographic direction). In the case of strain applied along one of the main crystal axes ([100], [010], [110], [-110]) the respective easy direction $\delta$ coincides with the strain direction. However, this does not have to be the case for a strain applied in a general direction[46] and, therefore, we treat $\delta$ as an unknown parameter.

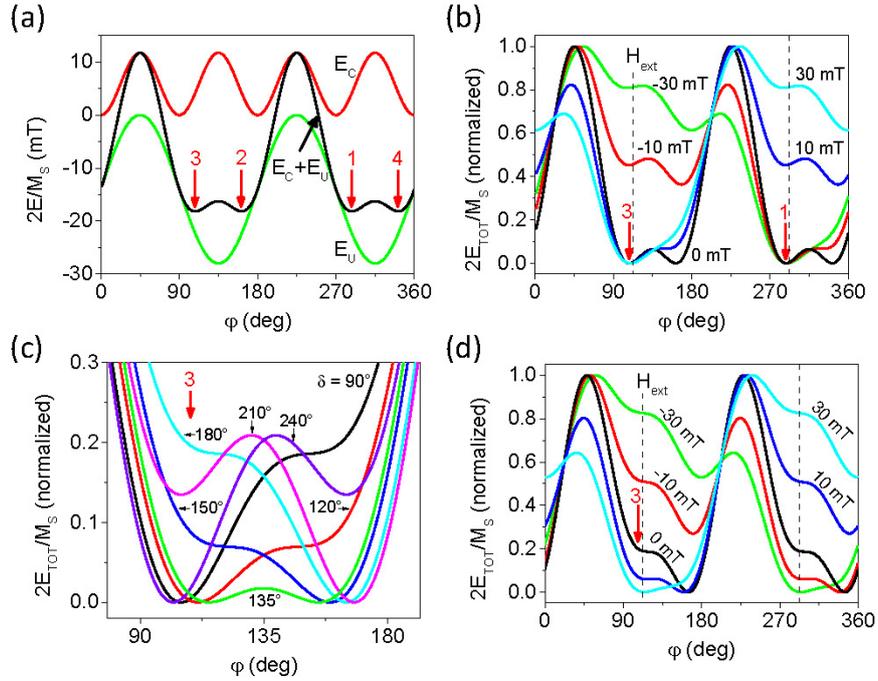

**FIG. 6. Magnetic energy density in unstrained and strained GaMnAs sample**. (a) Angular dependence of cubic ($E_C$), uniaxial ($E_U$) and total ($E_C + E_U$) magnetic anisotropy in a bare (unstrained) $Ga_{1-x}Mn_xAs$ epilayer with $x \approx 0.038$ where $\mu_0 H_C = 47$ mT and $\mu_0 H_U = 28$ mT. The positions of easy axes no. 1 - 4 shown in Fig. 4(a) are depicted by vertical arrows. (b) Change of magnetic anisotropy shown in (a) by $H_{EXT}$ applied along the direction depicted by a vertical dashed line both for positive ($\varphi_H = 112°$) and negative ($\varphi_H = 292°$) fields. (c) Change of magnetic anisotropy around EA 3 with a direction $\delta$ of the effective strain-induced uniaxial anisotropy of $\mu_0 H_S = 8$ mT for $\mu_0 H_{EXT} = 0$ mT. (d) Same as (b) for the case when also the strain-induced anisotropy with $\mu_0 H_S = 8$ mT and $\delta = 180°$ is included.

Our test $Ga_{1-x}Mn_xAs$ / piezo-actuator stack contains 20 nm thick epilayer with $x \approx 0.038$ where $\mu_0H_C$ = 47 mT and $\mu_0H_U$ = 28 mT.[38] In Fig. 6(a) we show the angular dependence of the resulting magnetic anisotropy that can be characterized by four EAs at 108° (EA 3), 161° (EA 2), 288° (EA 1), and 341° (EA 4). If an external magnetic field $H_{EXT}$ is applied in direction $\varphi_H$ that is close to the position of one EA (in our case $\varphi_H$ = 112° is close to EA 3 located at 108°), the magnetic anisotropy will change with $H_{EXT}$ as depicted in Fig. 6(b). In particular, if the magnetization is rotated to the direction corresponding to EA 3 by an application of large positive $H_{EXT}$, it will remain in this state until sufficiently strong $H_{EXT}$ of *opposite polarity* is applied, when the magnetization switching to the opposite magnetic state (EA 1) occurs. We note that this simple model of magnetization switching does not take into account a presence of defects which can serve as low-energy nucleation centers for the magnetization reversal process. Once the reversed domain is nucleated, the switching process can continue via a domain wall motion that will decrease the value of $H_{EXT}$ required to achieve the magnetization switching.

In a presence of mechanical strain – induced, e.g., by a piezostressor – the magnetic field-induced magnetization switching can be considerably different. In Fig. 6(c) we show the changes of the magnetic anisotropy around EA 3 with a direction $\delta$ of the effective strain-induced uniaxial anisotropy (with a magnitude $\mu_0H_S$ = 8 mT). In Fig. 6(d) we demonstrate that for $\delta$ = 180°, for example, the switching from EA 3 can occur even *before* the polarity of $H_{EXT}$ is changed, which is consistent with the experimental data shown in Fig. 5(a). However, as there exist several possible combinations of $H_S$ and $\delta$ that predict a rather similar behavior with $H_{EXT}$, Fig. 6(d) should be regarded solely as a qualitative explanation of the data in Fig. 5(a).

# REFERENCES

[1] M.M. Waldrop, Nature News 530, 144 (2016).

[2] *Spin Electronics*, edited by M. Ziese and M. J. Thornton (Springer, Berlin, 2001).

[3] *Semiconductor Spintronics and Quantum Computation*, edited by D. D. Awschalom, D. Loss, N. Samarth, series *NanoScience and Technology* (Springer, Berlin, 2002).

[4] J. Ferre, and G. A. Gehring, Rep. Prog. Phys. **47**, 513 (1984).

[5] H. Ebert, Rep. Prog. Phys. **59**, 1665 (1996).

[6] A. K. Zvezdin, V. A. Kotov, *Modern Magnetooptics and Magnetooptical Materials* (Institute of Physics Publishing, Bristol, 1997).

[7] M.I. Dyakonov and V.I. Perel in *Optical orientation*, edited by F. Meier and B. Zakharchenya Vol. 8 of *Modern problems in condensed matter sciences*, Chap. 2 (North-Holland, Amsterdam, 1984).

[8] J. Wunderlich, B. G. Park, A. C. Irvine, L. P. Zarbo, E. Rozkotová, P. Němec, V. Novák, J. Sinova, T. Jungwirth, Science 330, 1801 (2010).

[9] A. Hubert and R. Schäfer, *Magnetic Domains: The Analysis of Magnetic Microstructures* (Springer, Berlin, 1998).

[10] R. Schäfer, *Investigation of Domains and Dynamics of Domain Walls by the Magneto-optical Kerr-effect*, in *Handbook of Magnetism and Advanced Magnetic Materials*, edited by H. Kronmüller and S. Parkin, vol. 3. (John Wiley & Sons, 2007).

[11] J. McCord, J. Phys. D: Appl. Phys. **48**, 333001 (2015).

[12] Y. Acremann, M. Buess, C. H. Back, M. Dumm, G. Bayreuther, and D. Pescia, Nature **414**, 51 (2001).

[13] A. Kirilyuk, A. V. Kimel, and Th. Rasing, Rev. Mod. Phys. **82**, 2731 (2010).

[14] B. A. Ruzicka, L. K. Werake, H. Samassekou, and H. Zhao, Appl. Phys. Lett. **97**, 262119 (2010).

[15] H. Mino, S. Yonaiyama, K. Ohto, and R. Akimoto, Appl. Phys. Lett. **99**, 161901 (2011).

[16] N. Kumar, B. A. Ruzicka, N. P. Butch, P. Syers, K. Kirshenbaum, J. Paglione, and H. Zhao, Phys. Rev. B **83**, 235306 (2011).

[17] J. Li, M.-S. Lee, W. He, B. Redeker, A. Remhof, E. Amaladass, Ch. Hassel, and T. Eimüller, Rev. Sci. Instr. **80**, 073703 (2009).

[18] T. Henn, T. Kiessling, W. Ossau, L. W. Molenkamp, K. Biermann, and P. V. Santos, Rev. Sci. Instr. **84**, 123903 (2013).

[19] O. Schmitt, D. Steil, S. Alebrand, F. Ganss, M. Hehn, S. Mangin, M. Albrecht, S. Mathias, M. Cinchetti, and M. Aeschlimann, Eur. Phys. J. B **87**, 219 (2014).

[20] S. A. Crooker and D. L. Smith, Phys. Rev. Lett. **94**, 236601 (2005).

[21] A. W. Rushforth, E. De Ranieri, J. Zemen, J. Wunderlich, K. W. Edmonds, C. S. King, E. Ahmad, R. P. Campion, C. T. Foxon, B. L. Gallagher, K. Výborný, J. Kučera, and T. Jungwirth, Phys. Rev. B **78**, 085314 (2008).

[22] P. Němec, E. Rozkotová, N. Tesařová, F. Trojánek, E. De Ranieri, K. Olejník, J. Zemen, V. Novák, M. Cukr, P. Malý, and T. Jungwirth, Nature Physics **8**, 411 (2012).

[23] E. De Ranieri, P. E. Roy, D. Fang, E. K. Vehsthedt, A. C. Irvine, D. Heiss, A. Casiraghi, R. P. Campion, B. L. Gallagher, T. Jungwirth, and J. Wunderlich, Nature Materials **12**, 808 (2013).

[24] Y. Kato, R. C. Myers, A. C. Gossard, and D. D. Awschalom, Nature **427**, 50 (2004).

[25] D. Butkovičová, X. Marti, V. Saidl, E. Schmoranzerová-Rozkotová, P. Wadley, V. Holý, and P. Němec, Rev. Sci. Instrum. **84**, 103902 (2013).

[26] To match the maximum of the MO spectrum of the sample shown in Fig. 2(d), the central laser wavelength of 774 nm and filters F1(Thorlabs, NF785-33) and F2 (Thorlabs, FBH780-10), were chosen. Other combinations of central laser wavelength and filters can be used to cover different spectral regions. The filtering of the original laser pulse spectral content leads to a prolongation of the laser pulses. To address this, we have performed the intensity autocorrelation measurements both without and with the filters. We observed that the insertion of the filters leads to a prolongation of the autocorrelation trace from 150 to 320 fs (full width at half maximum), which corresponds to the temporal resolution of our pump-probe experiments.

[27] L. Nádvorník, P. Němec, T. Janda, K. Olejník, V. Novák, V. Skoromets, H. Němec, P. Kužel, F. Trojánek, T. Jungwirth, and J. Wunderlich, Scientific Reports **6**, 22901 (2016).

[28] S. Wiebel, J.-P. Jamet, N. Vernier, A. Mougin, J. Ferré, V. Baltz, B. Rodmacq, and B. Dieny, Appl. Phys. Lett. **86**, 142502 (2005).

[29] W. Rave, R. Schäfer, and A. Hubert, J. Magn. Magn. Mater. **65**, 7 (1987).

[30] T. von Hofe, N. O. Urs, B. Mozooni, T. Jansen, C. Kirchhof, D. E. Bürgler, E. Quandt, and J. McCord, Appl. Phys. Lett. **103**, 142410 (2013).

[31] N. Tesařová, T. Ostatnický, V. Novák, K. Olejník, J. Šubrt, H. Reichlová, C. T. Ellis, A. Mukherjee, J. Lee, G. M. Sipahi, J. Sinova, J. Hamrle, T. Jungwirth, P. Němec, J. Cerne, and K. Výborný, Phys. Rev. B **89**, 085203 (2014).

[32] P. Němec, M. Fiebig, T. Kampfrath, and A. V. Kimel, Nature Physics, submitted, arXiv: 1705.10600.

[33] S. Inoué and W. L. Hyde, J. Biophys. Biochem. Cytol. **3**(6), 831 (1957).

[34] In an ideal case, without the presence of the depolarization effects, this angle should be equal to the actual polarization rotation caused by the corresponding MO effect, which is typically of the order of

1 mrad. The depolarization generally leads to decrease of $C_{MO}$ and to an increase of the optimal analyzer opening angle. For further details see Ref. 33.

[35] T. Dietl, H. Ohno, Rev. Mod. Phys. **86**, 187 (2014).

[36] T. Jungwirth, J. Wunderlich, V. Novák, K. Olejník, B. L. Gallagher, R. P. Campion, K. W. Edmonds, A. W. Rushforth, A. J. Ferguson, and P. Němec, Rev. Mod. Phys. **86**, 855 (2014).

[37] N. Tesařová, J. Šubrt, P. Malý, P. Němec, C. T. Ellis, A. Mukherjee, and J. Cerne, Rev. Sci. Instrum. **83**, 123108 (2012).

[38] P. Němec, V. Novák, N. Tesařová, E. Rozkotová, H. Reichlová, D. Butkovičová, F. Trojánek, K. Olejník, P. Malý, R. P. Campion, B. L. Gallagher, Jairo Sinova, and T. Jungwirth, Nature Communications **4**, 1422 (2013).

[39] G. V. Astakhov, A. V. Kimel, G. M. Schott, A. A. Tsvetkov, A. Kirilyuk, D. R. Yakovlev, G. Karczewski, W. Ossau, G. Schmidt, L. W. Molenkamp, and Th. Rasing, Appl. Phys. Lett. **86**, 152506 (2005).

[40] A. A. Sapozhnik, R. Abrudan, Yu. Skourski, M. Jourdan, H. Zabel, M. Kläui, and H.-J. Elmers, Phys. Status Solidi RRL **11**, 1600438 (2017).

[41] M. Shayegan, K. Karrai, Y. P. Shkolnikov, K. Vakili, E. P. De Poortere, and S. Manus, Appl. Phys. Lett. **83**, 5235 (2003).

[42] M. D. Stiles in *Ultrathin magnetic structures III, Fundamentals of Nanomagnetism* edited by J. A. C. Bland, and B. Heinrich, Springer, Berlin, 2005, Ch. 4.

[43] T. Jungwirth, X. Marti, P. Wadley, and J. Wunderlich, Nature Nanotechnology **11**, 231 (2016).

[44] V. Saidl, P. Němec, P. Wadley, V. Hills, R. P. Campion, V. Novák, K. W. Edmonds, F. Maccherozzi, S. S. Dhesi, B. L. Gallagher, F. Trojánek, J. Kuneš, J. Železný, P. Malý, and T. Jungwirth, Nature Photonics **11**, 91 (2017).

[45] P. Wadley *et al.*, Science **351**, 587 (2016).

[46] J. Zemen, J. Kučera, K. Olejník, and T. Jungwirth, Phys. Rev. B **80**, 155203 (2009).